\documentclass[sigconf]{acmart}
\usepackage{listings}

\AtBeginDocument{%
  }

\setcopyright{acmlicensed}
\copyrightyear{2018}
\acmYear{2018}
\acmDOI{XXXXXXX.XXXXXXX}
\usepackage{rotating}
\usepackage{algorithm}
\usepackage{algorithmic}
\usepackage{comment} 
\usepackage{geometry}
\usepackage{array}
\usepackage{tabularx}

\lstdefinelanguage{MyLanguage}{
  morekeywords={begin, end, solve},
  morecomment=[l]{//},
  morestring=[b]",
}

\usepackage{color}  

\definecolor{codegreen}{rgb}{0,0.6,0}
\definecolor{codegray}{rgb}{0.5,0.5,0.5}
\definecolor{codepurple}{rgb}{0.58,0,0.82}
\definecolor{backcolour}{rgb}{0.95,0.95,0.92}

\lstdefinestyle{mystyle}{
    backgroundcolor=\color{backcolour},
    commentstyle=\color{codegreen},
    keywordstyle=\color{magenta},
    numberstyle=\tiny\color{codegray},
    stringstyle=\color{codepurple},
    basicstyle=\ttfamily\footnotesize,
    breakatwhitespace=false,
    breaklines=true,
    captionpos=b,
    keepspaces=true,
    numbers=left,
    numbersep=5pt,
    showspaces=false,
    showstringspaces=false,
    showtabs=false,
    tabsize=2
}

\usepackage[utf8]{inputenc}
\usepackage{hyperref}  
\usepackage{pdflscape}
\usepackage{booktabs}  
\usepackage{array}     
\usepackage{rotating}  
\usepackage{fontawesome5}
\usepackage{threeparttable}
\usepackage{tablefootnote}
\usepackage[most]{tcolorbox}  

\lstdefinelanguage{JavaScript}{
  keywords={typeof, new, true, false, catch, function, return, null, catch, switch, var, if, in, while, do, else, case, break},
  keywordstyle=\color{blue}\bfseries,
  ndkeywords={class, export, boolean, throw, implements, import, this},
  ndkeywordstyle=\color{darkgray}\bfseries,
  identifierstyle=\color{black},
  sensitive=false,
  comment=[l]{//},
  morecomment=[s]{/*}{*/},
  commentstyle=\color{purple}\ttfamily,
  stringstyle=\color{red}\ttfamily,
  morestring=[b]',
  morestring=[b]"
}

\newtcolorbox{takeawaybox}[1]{
    enhanced,
    colback=white,
    colframe=gray!30!white,
    coltitle=black,
    fonttitle=\bfseries,
    fontupper=\small,
    title={\large\faBook~#1},
    boxrule=1pt,
    sharp corners,
    boxed title style={
        colframe=white, 
        colback=gray!30!white, 
        sharp corners,
        left=1mm,
        right=1mm,
    },
    attach boxed title to top left={
        yshift=-\tcboxedtitleheight/2,
        xshift=1mm,
        yshifttext=-\tcboxedtitleheight/2,
    },
    top=0mm,
    bottom=0mm,
    left=3mm,
}

\usepackage{tikz}
\newcommand{\blackcircle}[1]{%
    \begin{tikzpicture}[baseline=-0.75ex]
        \node[circle, fill=black, inner sep=1.5pt, text=white, text width=0.5em, align=center] {#1};
    \end{tikzpicture}%
}
\usepackage{comment}

\acmISBN{978-1-4503-XXXX-X/18/06}

\begin{document}

\title{Experimenting with Multi-Agent Software Development: Towards a Unified Platform}

\author{Malik Abdul Sami}
\authornotemark[1]
\email{malik.sami@tui.fi}

\affiliation{%
  \institution{Tampere University}
  \city{Tampere}
  \country{Finland}}

\author{Muhammad Waseem Awan}
\authornotemark[1]
\email{muhammad.m.waseem@jyu.fi}

\affiliation{%
  \institution{Jyväskylä University}
  \city{Jyväskylä}
  \country{Finland}
}

\author{Zeeshan Rasheed}
\affiliation{%
  \institution{Tampere University}
  \city{Tampere}
  \country{Finland}}
\email{zeeshan.rasheed@tuni.fi}

\author{Mika Saari}
\affiliation{%
  \institution{Tampere University}
  \city{Tampere}
  \country{Finland}
  \
\email{mika.saari@tuni.fi}  
}

\author{Kari Systä}
\affiliation{%
  \institution{Tampere University}
  \city{Tampere}
  \country{Finland}}
\email{kari.systa@tuni.fi}

\author{Pekka Abrahamsson}
\affiliation{%
  \institution{Tampere University}
  \city{Tampere}
  \country{Finland}}
\email{pekka.abrahamsson@tuni.fi}




\begin{abstract}
  Large language models are redefining software engineering by implementing AI-powered techniques throughout the whole software development process, including requirement gathering, software architecture, code generation, testing, and deployment. However, it is still difficult to develop a cohesive platform that consistently produces the best outcomes across all stages. The objective of this study is to develop a unified platform that utilizes multiple artificial intelligence agents to automate the process of transforming user requirements into well-organized deliverables. These deliverables include user stories, prioritization, and UML sequence diagrams, along with the modular approach to  APIs, unit tests, and end-to-end tests. Additionally, the platform will organize tasks, perform security and compliance, and suggest design patterns and improvements for non-functional requirements. We allow users to control and manage each phase according to their preferences. In addition, the platform provides security and compliance checks following European standards and proposes design optimizations. We use multiple models, such as GPT-3.5, GPT-4, and Llama3 to enable to generation of modular code as per user choice. The research also highlights the limitations and future research discussions to overall improve the software development life cycle. The source code for our uniform platform is hosted on GitHub, enabling additional experimentation and supporting both research and practical uses.
\end{abstract}




\maketitle

\section{Introduction}

The Software Development Life Cycle (SDLC) includes methodologies and practices that development teams employ to plan, architect, implement, test, deploy, and maintain software \cite{belzner2023large}.
The first phase,
Requirements engineering plays a crucial role in software engineering by linking technical specifications designed by engineers with the ultimate goals of software systems \cite{bjarnason2016multi}. During all stages of development, including proposal development, design, implementation, and testing, the team regularly reviews and refines these requirements \cite{9487986, ahmad2023towards}. This ongoing process ensures the software fulfills user needs, which are typically expressed as user stories or tasks. One of the main challenges is the accurate translation of user requirements into technical specifications that accurately meet users' expectations. Software architecture and design become essential once the team defines the requirements clearly and prioritizes them \cite{SAFe_WSJF, chang2024survey}. UML diagrams help developers understand how to implement user stories \cite{abrahamsson2017agile, mucha2024systematic}. This phase emphasizes the importance of meeting requirements and keeping track of changes or mismatches during development.

The development phase in the SDLC focuses on writing code based on user requirements. It is important to maintain thorough documentation to help developers write backend, frontend, or database code. Testing, which includes both unit and end-to-end testing, is another key step in the project. As the need for faster development grows, the field of software engineering is rapidly evolving. It is critical to identify precise and clear requirements necessary for using a test case scenarios suite that effectively validates and verifies software outputs \cite{tang2024chatgpt, svensson2024not}.

Artificial Intelligence (AI) and Natural Language Processing (NLP)
enable new generation
of computational tools that can generate text with near-human accuracy using generative AI \cite{ebert2023generative}. The development of advanced language models, particularly within the Generative Pre-trained Transformer (GPT) series, represents a significant advancement in this area \cite{radford2018improving, rasheed2024can, ouyang2022training}.  The recent rise of  Large Language Models (LLMs) has started a transformative time in programming and computer program building \cite{xie2023chatunitest}. Trained on broad datasets related to coding, these LLMs have created a careful understanding of programming, which permits them to exceed expectations in different code-related errands. Examples of such tools include LLaMA and ChatGPT \cite{touvron2023llama, roumeliotis2023chatgpt}.

These LLMs are good at understanding natural language, marking a major shift in software engineering from simple tools to collaborative aids. LLMs provide insights and analysis that were not possible with traditional methods \cite{sami2024system, rasheed2023autonomous, macneil2023experiences}. Recently, multi-agent systems have made significant progress in solving complex problems in software development by emulating development roles \cite{rasheed2024codepori, samiprioritizing, feng2024prompting, waseem2023artificial}. Lin et al. \cite{lin2024llm} paper introduces a code generation framework inspired by established software engineering practices. Leverages multiple LLM agents to emulate various software process models, namely LCGWaterfall, LCGTDD, and LCGScrum \cite{lin2024llm}. Tang et al. \cite{tang2024collaborative} developed self-collaboration, which assigns LLM agents to work in collaborative software development. Wang et al.\cite{wang2024xuat} proposed an LLM-powered multi-agent collaborative system, named XUAT-Copilot, for automated UAT . While some research has explored integrating multi-agents within LLM frameworks \cite{qian2023communicative, waseem2023using, zhou2023language}, their research focus diverges from the influence of the software process model on software design and code generation of SDLC.

Despite the promising applications of LLMs in automating software engineering tasks, it is important to recognize that software development is a collaborative and teamwork effort \cite{kirova2024software}. In the real world, developers and stakeholders work together, by following a unified platform that meets all the stakeholders under one umbrella \cite{rinard2024software}. They improve the process but do not provide a unified platform approach where requirements to user stories, prioritization, design architecture, and both generations of frontend and end-to-end with security, best practices, and patterns according to standards are defined \cite{quevedo2024evaluating}.

Using LLMs to convert requirements into user stories, and then to design, generate code, and generate test cases is a necessary step in the ever-changing software development industry. This approach meets the critical need to automate, enhance, and perfect the creation and validation of requirements through acceptance criteria \cite{tikayat2023agile, zhang2024llm, li2024sheetcopilot}.

The \textbf{objective} of this research is to develop a comprehensive software development platform that utilizes various LLM-based agents. This platform is designed to convert user requirements into structured software development outputs such as user stories, prioritization, UML diagrams, APIs, unit tests, and end-to-end tests. It will generate code in Python and use frameworks suitable for both backend and frontend components. Additionally, the platform will organize tasks, perform security and compliance, and suggest design patterns and improvements for non-functional requirements. The ultimate goal is to enhance the efficiency, compliance, and quality of software development by integrating generative AI and human involvement in each phase of the software development lifecycle, ensuring the software is ready for deployment. 

The \textbf{key contributions} of this study are summarized as follows:

\begin{itemize}
    \item We enhance our existing work by incorporating additional prioritization techniques into a unified platform that generates user stories and prioritizes requirements using large language models (LLMs).
    \item We have bridged the gap
    between requirements engineering and technical development
    by generating UML diagrams with the help of the LLM agent and PlantUML for the software Architect phase.
    \item We generate a modular
    Python code with unit tests,
    and for the client side using React and end-to-end test cases.
    \item We integrate a multi-AI agent model that utilizes LLMs to automate the SDLC process with user involvement, significantly speeding up the SDLC using GPT-3.5, GPT-4, and llama3.
\end{itemize}


\textbf{Structure of the Paper}: Section \ref{Background} reports the background work releated to this study. Section \ref{Methodology} explains the proposed unified platform. Section \ref{Results} presents the initial results, and Section \ref{Discussion} discusses the key findings. Section \ref{ThreatstoValidity} discusses the possible threats and limitations of this study. Section \ref{Conclusions} concludes the study.


\section{Background}
\label{Background}
In this section, we briefly present the background of the study with a focus on existing research. Section \ref{Generative AI} provides an overview of studies concerning generative AI and its application across software engineering. Section \ref{LLM in Software Development} examines works that have utilized LLMs for software development.

\subsection{Generative AI in Software Engineering}
\label{Generative AI}

Generative AI assists in creating outputs that mimic human performance across diverse fields including multimedia production, computer vision, and NLP \cite{hacker2023regulating}. The goal is to generate new content. Using autoregressive language models such as GPT and Generative Adversarial Networks (GANs), this technology transforms text production, machine translation, conversational systems, and code generation \cite{aydin2023chatgpt, waseem2023using}. The transformer architecture, which significantly enhances NLP capabilities by recognizing contextual relationships in texts, is crucial for the GPT model's ability to produce text that closely resembles human writing \cite{radford2018improving}.

Recent advancements show the versatility and effectiveness of GPT models and their potential to revolutionize software engineering methods \cite{liu2023gpt,ouyang2022training}. These models enable the automation of several tasks in the SDLC, such as error identification, code snippet creation, and documentation, by utilizing large code databases for training \cite{feng2023investigating,treude2023navigating}. Integrating GPT models into SE enhances the efficiency and quality of software production, and accelerates coding and application development \cite{dong2023self,ma2023scope}.

The emergence of generative AI in SE marks a significant shift, emphasizing its capability to refine and speed up software development processes. Integrating GPT models into SE heralds a fundamental transformation, impacting how software is crafted and maintained. Importantly, this advancement streamlines the prioritization of requirements—a key phase in development. The application of AI allows developers to generate test suites and test case scenarios more effectively, ensuring that critical features align with user needs and organizational objectives. This approach promises a move towards a more user-centric, and efficient paradigm in software development and testing \cite{wang2024software}.

\subsection{Large Language Model in Software Development}
\label{LLM in Software Development}
Code generation stands as a crucial research area because it helps reduce development costs. Training more powerful LLMs and exploring prompt-based and agent-based techniques for code generation are currently the most promising research directions.

Agent-based code generation focuses on the importance of defining roles and facilitating communication among multiple LLM agents. Some methods use external tools as agents. One such example is the test executor agent introduced by Huang et al. \cite{huang2023agentcoder}, which provides test logs for LLMs via a Python interpreter. A debugger agent that generates control flow graph information using a static analysis tool is introduced by Zhong et al. \cite{zhong2024ldb} to assist LLMs in finding flaws. Several other studies designate LLMs as agents by simulating human jobs, including chief technology officers, analysts, engineers, testers, and project managers. To tackle difficult and innovative challenges in autonomous software development, Zeeshan et al. \cite{rasheed2024codepori} present a codePori that uses LLM-based multi-AI agents. For example, system design, code development, code review, code verification, and test engineering are all tasks that each agent works on specifically. 

Many studies employ GPT models across the SDLC, including requirements gathering, code generation, unit, and GUI testing \cite{rajbhoj2024accelerating}. LLMs are used to develop a unified platform that potentially makes software development more efficient and effective. However, challenges such as the generation of inaccurate information (e.g., hallucinations) and limitations in natural language comprehension and domain knowledge highlight a significant need for further research \cite{sauvola2024future}. A solution is necessary that can understand diverse client needs, automatically generate user stories, create UML diagrams, and use LLMs to efficiently organize backend and end-to-end test cases \cite{nguyen2023generative, sami2024prioritizing}. Modern software development methods that prioritize flexibility and intelligent use of LLM agents should be adopted by this tool. By meeting this need, we can improve software quality and fully leverage AI in the SDLC. This approach opens up substantial opportunities for new research and development in the field by integrating an LLM-supported web-based platform, ensuring all critical scenarios are considered and meet user needs.

\section{Proposed Multi-Agent Unified Platform}
\label{Methodology}
\begin{figure*}[t]
  \centering
  \includegraphics[width=\textwidth]{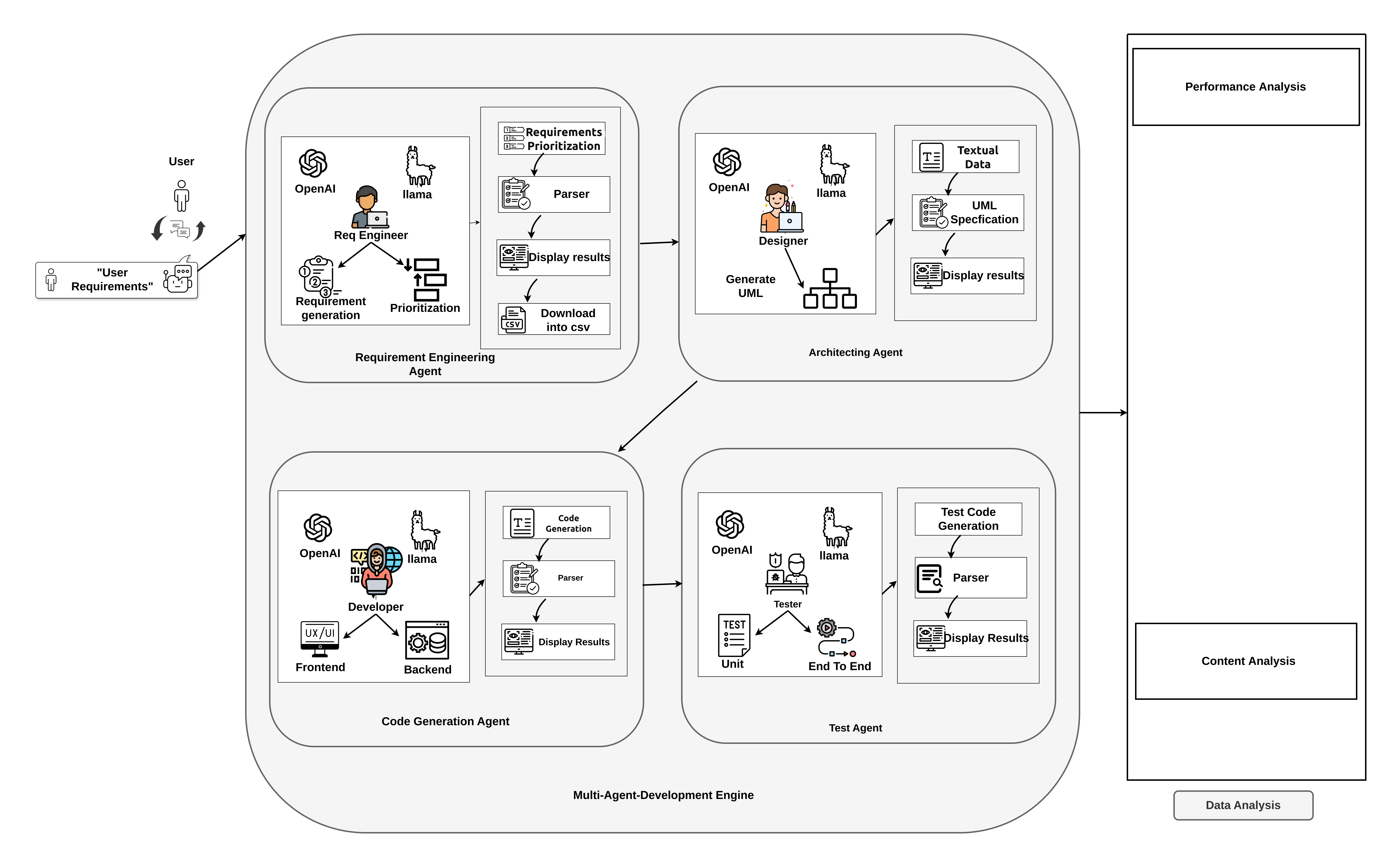}

  \caption{Developing a unified software development platform through multi-agent collaboration}
  \label{fig:unified_framework}
  \Description{Multi agents collaboration to build a unified framework for SDLC}
\end{figure*}

The architecture of our LLM-based multi-agent model is specifically designed to streamline and enhance the SDLC by incorporating multiple specialized AI agents \cite{rasheed2024large}. Each agent is responsible for handling different aspects of the development process, from requirements engineering to testing. This section outlines the functions and processes of these agents as shown in the provided Figure \ref{fig:unified_framework}.

\subsection{Prompt Engineering}

Before exploring each section, we discuss the role of prompt engineering, which is employed in every area of the SDLC. It is considered crucial for generating content. The better the prompt quality, the better the expected results. Below, we have provided the tasks of each agent and their respective work in Table \ref{Table 01}.

\begin{table*}[ht]
\centering
\caption{Prompts and tasks for various agents.}
\label{Table 01}
\small 
\setlength{\tabcolsep}{3pt} 
\renewcommand{\arraystretch}{1.2} 
\begin{tabular}{|l|l|p{3.6cm}|p{10cm}|}
\hline
\textbf{No} & \textbf{Agent} & \textbf{Main Work} & \textbf{Prompt Content} \\ \hline
1 & RE Agent-01 & User Story Generation & Generate user stories from the requirments. \\ \hline
2 & RE Agent-02 & Prioritization & Prioritized user stories Using AHP, WSJF techniques \cite{sami2024prioritizing}. \\ \hline
3 & Architecting Agent & UML Conversion & Act as a senior architect to create PlantUML code from project descriptions, structuring responses with Code and Critique sections. \\ \hline
4 & Code Generation Agent & Python and React Code Generation & Act as a senior developer for React and Python, generating clean, optimized code based on project requirements. Structured responses include Response, Code . \\ \hline
5 & Test Agent & API and Frontend Testing & Generate unit tests for API and end-to-end tests for frontend applications, ensuring coverage and effectiveness. \\ \hline
6 & Compliance Agent & European Standards Compliance & Ensure adherence to European rules, regulations, and security standards, focusing on compliance checks and documentation. \\ \hline
\end{tabular}
\end{table*}

\subsection{Requirements Engineering Agent}

The Requirements Engineering agent is an integral part of our engine, leveraging AI capabilities to automate the generation and prioritization of software requirements \cite{sami2024prioritizing}. This agent begins its process by collecting user requirements. Utilizing APIs from OpenAI and Llama, it interprets and translates these user inputs into structured software requirements. Once generated, the agent employs prioritization algorithms to order these requirements based on factors such as importance, feasibility, and potential impact. The prioritized requirements are then displayed to the requirement engineer and can be downloaded in CSV format for further analysis or documentation purposes.

\subsection{Architecting Agent}

The Architectural agent assumes responsibility for converting the prioritized requirements into useable design artifacts after requirements engineering. The creation of Unified Modeling Language (UML) diagrams, which provide as a visual depiction of the system architecture and interactions, is the exclusive emphasis of this agent. The agent passes the textual requirements to PlantUML \cite{PlantUML} after parsing the structured requirements to determine the required components and their relationships using OpenAI and llama APIs \cite{metallama}. System designers can then examine and improve the generated UML diagrams, making sure that all system requirements are sufficiently documented during the design phase.

\subsection{Code Generation Agent}
The Code agent uses AI to automate the coding process throughout the code generation phase, improving frontend and backend development. This agent converts system designs from UML diagrams into executable code with the help of llama API keys and OpenAI. It ensures that the program is both functional and aesthetically pleasing by concentrating on user interface and user experience components in front-end development. For backend development, the agent creates the server-side code needed to manage data, implement application logic, and communicate with databases and other services. After that, the resulting code is shown to the programmers for approval and integration into the overall system.

\subsection{Testing Agent}

Our engine's testing agent, which is in charge of thoroughly evaluating the software to check its dependability and quality, is the last stage. Unit tests as well as end-to-end tests are generated and run by this app. While end-to-end tests evaluate how different system components interact with one another under simulated real-world settings, unit tests are used to confirm the functionality of particular components. The Testing agent uses llama APIs and OpenAI to automate the creation and execution of test cases. It then presents the findings to testers so they can see any problems and fix them before the program is released.

\subsection{Data Analysis}

The section includes a strong data analysis to measure the overall effectiveness of the suggested modules. To evaluate the multi-agent system's efficacy in various projects and scenarios, this module has to analyze, how the system is built and what is the performance of the APIs of each life cycle. We can consistently improve our AI models and development procedures by assessing these factors, which will vow the best possible performance and dependability from the software we produce.

\subsection{Technical Background}
In this section, we provide the technical details of our proposed unified platform. We proposed 3 algorithms to utilize LLM for SDLC along with the code available on Github \cite{GPTLaboratory2024}. Initially, \textbf{Algorithm 01} collected the requirements, which can be supplied either through direct user input or via files uploaded from the client side. Once the data is entered into the system, the algorithm utilizing an OpenAI-powered LLM—processes and analyzes the requirements. The ultimate output of the algorithm is a set of prioritized user stories, structured to reflect their importance and urgency within the project scope. These user stories are essential for subsequent phases of the SDLC, such as planning and implementation. Additionally, the algorithm provides an option to export these prioritized user stories in a CSV format, facilitating easy integration with other project management tools and ensuring a seamless transition to later stages of the project development.

\begin{algorithm}
\hspace*{\algorithmicindent} \textbf{Input}: Requirements $data$ provided via user input or client-side upload. \\
\hspace*{\algorithmicindent} \textbf{Output}: Prioritized user stories with an option to download as CSV.

\caption{Requirements Engineering Process with LLM-based agents}
\label{algo:req_engineering_comprehensive}

\begin{algorithmic}[1]
\STATE \textbf{Collect Requirements:}
\STATE User inputs requirements or uploads from the client side (ReactJS).
\STATE \textbf{Generate User Stories:}
\STATE $objective \gets$ Extract objective from $data$
\STATE $num\_stories \gets$ Determine number of stories needed
\STATE $user\_stories \gets$ generate\_user\_stories\_with\_epics
\STATE \textbf{Prioritize User Stories:}
\FOR{each $method$ in \{AHP, MoSCoW, 100 Dollar\}}
    \IF{$method$ is AHP}
        \STATE $prompt \gets$ construct\_ahp\_prompt($user\_stories$)
    \ELSIF{$method$ is MoSCoW}
        \STATE $prompt \gets$ construct\_moscow\_prompt($user\_stories$, $method$)
    \ELSIF{$method$ is 100 Dollar}
        \STATE $prompt \gets$ construct\_100\_dollar\_prompt($user\_stories$)
    \ENDIF
    \STATE $prioritized\_stories \gets$ call\_openai\_api($prompt$)
    \STATE Parse and store $prioritized\_stories$ based on $method$.
\ENDFOR
\STATE \textbf{Download Prioritized Stories:}
\STATE Generate CSV of prioritized user stories based on final prioritization.
\STATE Provide a download link to the user through a client-side application.

\end{algorithmic}
\end{algorithm}

The process begins with the \textbf{Collect Requirements} step, where requirements are gathered either through direct user input or via file uploads from the client side, specifically using ReactJS. Once the requirements are collected, the algorithm transitions to the \textbf{Generate User Stories} phase. The main objective is extracted from the data, and the number of necessary user stories is determined. User stories, along with their corresponding epics, are then generated based on the extracted objectives.
Following the generation of user stories, Three distinct prioritization techniques are used in the \textbf{Prioritize User Stories} phase: the 100 Dollar Test, WSJF, and the Analytic Hierarchy Process (AHP). Appropriate prompts are created using the OpenAI API to make the prioritization process easier, depending on the approach chosen. By using these prompts, it is made sure that user tales are assessed based on their urgency and significance. A refined collection of prioritized user stories that are parsed and saved appropriately is the result of each prioritizing technique.

Downloading a CSV file with the prioritized user tales is the last step, \textbf{Download Prioritized tales}. A client-side application is used to make this file available for download, allowing for easy integration into later stages of project management and development. This method not only automates a large amount of the planning and requirements phases but also makes sure that the prioritization of user stories is both organized and aligned with the project's goals.

\begin{algorithm}
\hspace*{\algorithmicindent} \textbf{Input}: User stories as input, selection of model and method (e.g., 'gpt-3.5-turbo', 'gpt-4', 'llama3', 'UML-diagram'). \\
\hspace*{\algorithmicindent} \textbf{Output}: Visual UML diagram based on the user stories, displayed on the client side.

\caption{Software Architecting Agent Process}
\label{algo:software_design_agent}

\begin{algorithmic}[1]
\STATE \textbf{Select Model and Method:}
\STATE Determine the model and method based on user input or system configuration.
\STATE \textbf{Prepare API Call:}
\STATE Open the template file corresponding to the method (e.g., "agentplantUML.txt") for the selected model.
\STATE Replace the placeholder text with an actual project description to form the bot prompt.
\STATE Add the bot prompt to the conversation history.

\STATE \textbf{Call OpenAI API:}
\STATE Configure API headers and post data.
\STATE Make the API call to the selected model endpoint with the configured data.
\STATE \textbf{Handle API Response:}
\IF{API call is successful}
    \STATE Extract text and code blocks from the API response.
    \STATE Separate UML code using specific markers (e.g., "@startuml", "@enduml").
\ELSE
    \STATE Raise an exception and output the error.
\ENDIF

\STATE \textbf{Generate UML Diagram:}
\STATE Call the PlantUML open-source API with the extracted UML code.
\STATE Fetch the UML diagram in byte format.
\STATE \textbf{Display UML Diagram:}
\STATE Send the UML diagram to the client-side application for display.
\STATE Optionally, handle errors and exceptions if the diagram generation fails.

\end{algorithmic}
\end{algorithm}

An automated system for producing UML diagrams based on user stories is described in \textbf{Algorithm 02}. The purpose of this technique is to improve the software design phase by automating the generation of architecture diagrams, which are essential for developer instruction and documentation. The first step of the procedure involves choosing the right language model and technique depending on predetermined system parameters or user input. Methods like \textit{UML-diagram} and models like 'get-3.5-turbo', 'gpt-4', and 'llama3' are among the options. After making a decision, the system opens the template file for the selected method to get ready to access the OpenAI API. The project description is then added to this template, which already has placeholder content, to create a thorough bot prompt that is added to the conversation history.
An API request is performed to the chosen model endpoint with headers and data that have been defined specifically. The system collects text and code blocks, especially UML code indicated by certain markers like \textit{@startuml} and \textit{@enduml}, if the API request is successful. The system is built to raise an exception and offer an error output in the case of an API failure. After that, the \textit{PlantUML} open-source API receives the extracted UML code and creates a byte-formatted UML diagram. Ultimately, the client-side application receives this diagram and displays it, providing a clear and graphical depiction of the user narrative. This approach guarantees that complicated UML diagrams are correct representations of the architectural requirements of the project, while simultaneously making the process of creating them simpler.

The purpose of the \textbf{Algorithm 03} is to automate the process of creating source code that satisfies particular needs. This is accomplished by using a JSON object that is retrieved from an HTTP POST request. This object holds important information about the model that should be processed, the method to be used, and the content that needs to be generated into code.
First, the algorithm takes the requested data and extracts the three key parameters: content, model, and method. After that, the environment is configured using the model and method that have been defined. This entails configuring an authentication key, a related API URL, and a bot prompt file. Then, after loading the bot prompt file, placeholders in it are changed to reflect the material that was supplied in the input.

The system formats the post data to include the model and the customized bot prompt after the environment is suitably established. It then creates authentication headers using the key that was supplied. The required headers and data are then included in a POST request that is delivered to the given URL.
The program examines the response it receives from the API call to determine whether the request was successful. Code blocks that are particularly created depending on the input parameters, the chosen model, and the method are extracted and formatted from the payload if the process is successful. The answer then contains these code chunks, which can be used in testing or development settings. If the API call fails, the algorithm manages the issue and produces an appropriate error response. This robust process ensures that developers can obtain tailored code efficiently and reliably, enhancing productivity and facilitating specific development tasks.

\begin{algorithm}
\hspace*{\algorithmicindent} \textbf{Input}: JSON object from remote request containing `content`, `model`, and `method`. \\
\hspace*{\algorithmicindent} \textbf{Output}: Generated modular code for the specified method and section of development.
\caption{Modular Code Generator for Development and Testing}
\label{algo:modular_code_generator}

\begin{algorithmic}[1]
\STATE \textbf{Extract Input:} $content, model, method \gets$ Parse request data

\STATE \textbf{Configure Environment:}
\STATE $bot\_prompt\_file, url, key \gets$ Configure based on $model$ and $method$
\STATE $bot\_prompt \gets$ Load and replace placeholders in $bot\_prompt\_file$ with $content$

\STATE \textbf{Prepare Request:}
\STATE $headers \gets$ Create authentication headers using $key$
\STATE $post\_data \gets$ Format post data with $model$ and $bot\_prompt$

\STATE \textbf{API Call:}
\STATE $response \gets$ Send POST request to $url$ with $headers$ and $post\_data$
\STATE $status, payload \gets$ Process response checking for success

\STATE \textbf{Output Processing:}
\IF{$status = \text{"success"}$}
    \STATE $code\_blocks \gets$ Extract and format code from $payload$
    \STATE \textbf{return} Create response with $code\_blocks$
\ELSE
    \STATE \textbf{return} Handle error and create error response
\ENDIF
\end{algorithmic}
\end{algorithm}


\section{Preliminary Results}
\label{Results}
This section reports the preliminary results of the proposed LLM-based multi-agent platform. Our initial findings indicate that the proposed platform is capable of automating the whole process of SDLC with a modular approach. Below, we present the results of our proposed platform in Section \ref{Proposed Model} along with challenges and possible future agenda.

\label{Proposed Model}
\subsection{Requirement Engineering Phase}
We create user stories based on user needs in the requirements engineering area. We apply prioritization techniques like AHP (Analytic Hierarchy Process) and WSJF (Weighted Shortest Job First) in conjunction with LLMs (Large Language Models) to prioritize needs and recommend the best potential values. Algorithm 1 provides additional information about the specifics.

For generating requirements and prioritization, we explore the idea of automating systematic literature reviews (SLR) using large language models and generative AI \cite{sami2024system}. We present requirements and instruct agents to convert these requirements into user stories and epics. The input we pass into the platform can be seen below:

\begin{tcolorbox}
[colback=gray!5!white,colframe=gray!75!black,title=Snippet of User Requirements]
\scriptsize
`` \ldots~The proposed system is designed to streamline research processes by offering a suite of advanced features With a login feature. It enables users to generate research-specific questions, interact through a React-based user interface, and choose between ChatGPT 3.5 or 4 for language modeling. The system supports API integration, adheres to specific inclusion and exclusion criteria for literature review, and facilitates paper summarization and data extraction.
\end{tcolorbox}


This algorithm assists in managing software development processes by first collecting user requirements either through direct input or uploads via the client-side application. Once the requirements are collected, the algorithm extracts the main objectives and determines the necessary number of user stories, each associated with broader work themes called epics. Following this, it prioritizes these stories using various methods like AHP (Analytic Hierarchy Process), Moscow (Must have, Should have, Could have, Would like to have), or the 100 Dollar Test, depending on the chosen method. Each method generates a specific prompt that is processed by an AI API to rank the stories in order of importance. The results are then compiled into a CSV file, and a download link is provided for users through the frontend application, allowing them to access the prioritized user stories conveniently. This structured approach ensures efficient planning and execution in software development projects. Below we provided the result of prioritized work in Table \ref{Table02}.

\begin{table*}[ht]
\centering
\scriptsize
\caption{User Stories and Weighted Shortest Job First (WSJF) Prioritization Score [Business
Value (BV), Job Size (JS), Risk Reduction (RR), Time Criticality (TC)]}
\label{Table02}
\small 
\setlength{\tabcolsep}{3pt} 
\renewcommand{\arraystretch}{1.5} 
\begin{tabular}{|c|p{13cm}|c|c|c|c|c|}
\hline
\textbf{\#} & \textbf{User Stories} & \textbf{BV} & \textbf{JS} & \textbf{RR} & \textbf{TC} & \textbf{WSJF} \\

\hline
1 & As a research student, I want to be able to formulate research-specific questions within the system so that I can define the scope of my study effectively. & 7 & 5 & 6 & 4 & 3.4 \\
\hline
2 & As a researcher, I need a user-friendly React-based user interface to interact with the system seamlessly and navigate through various features effortlessly. & 8 & 4 & 5 & 6 & 4.75 \\
\hline
3 & As a language modeling enthusiast, I want the option to choose between ChatGPT 3.5 and 4 for my research tasks to leverage the latest advancements in natural language processing. & 6 & 6 & 8 & 7 & 3.5 \\
\hline
4 & As a researcher, I require API integration within the system to easily access external data sources and tools for my research projects. & 9 & 7 & 7 & 8 & 3.42 \\
\hline
5 & As a systematic reviewer, I aim to define precise inclusion and exclusion criteria for literature review within the system to ensure the relevance and quality of the research materials. & 8 & 5 & 6 & 5 & 3.8 \\
\hline
6 & As a busy researcher, I seek a feature that enables paper summarization and data extraction to save time and efficiently extract key information from research articles. & 7 & 6 & 7 & 6 & 3.33 \\
\hline
\end{tabular}
\end{table*}

\textit{Performance} is demonstrated by the AHP-based prioritization process completed in an average of six seconds of generating 10 stories and epic. Weighted Shortest Job First (WSJF) eight to nine seconds to suggest values and prioritize and display results in client-side both techniques showcasing the method's effectiveness and the system's capability to quicken the user story generation and prioritization phases.

In this section, we utilize LLMs to prioritize software requirements using various established methods, including the Analytic Hierarchy Process (AHP), Weighted Shortest Job First (WSJF), and the 100-dollar technique. To the best of our knowledge, this is the first investigation into the use of LLMs for prioritization within requirement engineering. Despite the promising results, several challenges must be addressed to enhance the robustness and applicability of these models. Below, we outline these challenges and propose a future research agenda in this emerging field.

\textbf{Challenges:} While we focus on the state-of-the-art applications of LLMs in requirement engineering, some practical challenges still need to be addressed:

A few of the main challenges facing requirement engineers are the necessity for accurate and unambiguous user stories with well-defined acceptance criteria, the unpredictability of results given the context, the frequent modifications of requirements, and inconsistent output with page refreshes. Subsequent paths are intended to improve requirement engineering efficacy in multiple ways. These include creating a co-pilot feature for automated requirement generation, expanding prioritization techniques to meet a variety of needs, improving document automation for various business documents using advanced AI techniques, and integrating open-source LLMs for cost-effectiveness and performance evaluation.

\begin{takeawaybox}{Takeaways}
\scriptsize
 \blackcircle{1} \textbf{Efficient Requirement Prioritization with LLMs}: The integration of Large Language Models with techniques like AHP and WSJF significantly speeds up and enhances the accuracy of requirement prioritization in software development.\\
 \blackcircle{2} \textbf{Associated Challenges and Future Directions}: The use of LLMs in requirement engineering faces challenges like consistency and accuracy, with future research aimed at expanding capabilities and improving the robustness of these technologies
\end{takeawaybox}

\subsection{Architecting Phase}

In the architecture section, we work on software UML diagrams using an AI agent and PlantUML. We use LLM to take requirements from the user and convert them into a specific textual format required by PlantUML, referred to as the PLANTUML response. We then pass this response to the client-side application using the MIME type svg/xml. The details are further explained in Algorithm 2.

\begin{figure}[h]
  \centering
  \includegraphics[width=\linewidth]{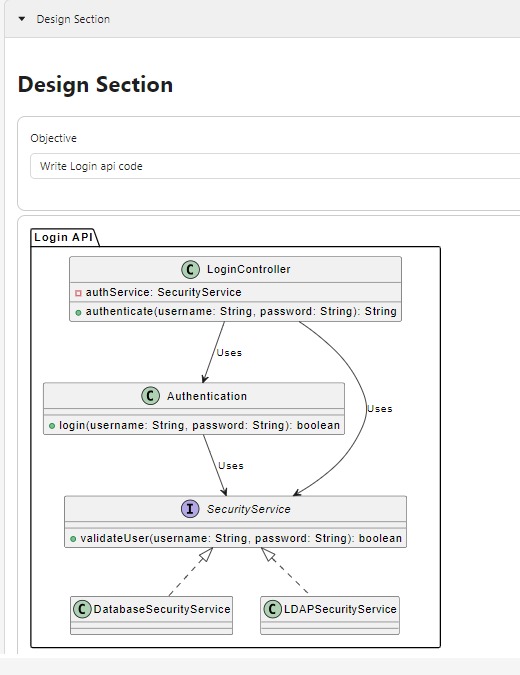}
  \caption{Login Code Design}
  \label{Figure 5}
  \Description{A UML diagram for the login code.}
\end{figure}

This algorithm describes the process of creating visual UML diagrams from user stories with the help of AI models like GPT-3.5 Turbo, GPT-4, or llama3. Initially, the model and method are selected based on user preferences or system settings. The algorithm prepares for an API call by opening a template file specific to the chosen method and filling it with the actual project description, creating a prompt for the AI. This prompt is then sent through an API call to the selected model's endpoint. If the API call is successful, the algorithm extracts UML code from the response, identifies the start and end of the UML block, and sends this code to the PlantUML API, which generates the UML diagram in a byte format. Finally, the diagram is sent to the client-side application for display. If there are any issues during the API call or diagram generation, the system raises an exception and reports the error. This process automates the creation of UML diagrams, ensuring that software architectural designs are accurately visualized based on the initial user stories. Below we provide the result in Figure \ref{Figure 5}.

In the architecture phase, we use LLM-based agents to convert into textual specifications, which draw a UML diagram however there are some areas where the challenges need to be addressed. 

Challenges with using large language models (LLMs) for generating UML specifications include occasional failures in accurate conversions, inconsistent results across different types of diagrams, and variability in outputs upon page refresh. To address these issues, future developments are focusing on integrating a variety of open-source LLMs to enable performance comparisons and broadening the range of supported UML diagrams. Additionally, there are plans to fine-tune these models specifically for creating detailed and accurate architect diagrams based on user requirements, enhancing their effectiveness and reliability in software development.




    

\begin{takeawaybox}{Takeaways}
\scriptsize
 \blackcircle{3} \textbf{AI-Powered UML Diagram Creation}: Utilizing LLM-based agents, the architect process automates the transformation of user stories into textual specifications for PlantUML, enabling the swift generation of accurate UML diagrams displayed via client-side applications.\\
 \blackcircle{4} \textbf{Associated Challenges and Future Directions}: The process faces challenges with occasional failures in accurate UML specification conversion and diagram consistency, prompting future enhancements including exploring diverse open-source LLMs and expanding UML diagram types and model training for optimized architect outputs.
\end{takeawaybox}

\subsection{Code Generation and Test Phase}

In this section, we explore four subsections: one for backend code generation, one for frontend code generation, one for generating unit tests for backend APIs, and one for generating test cases for the client side. Our contribution involves generating code for each subsection, including frontend, backend unit tests, or end-to-end scenarios.

The Code Generator for Development and Testing algorithm processes code generation requests efficiently. It starts by receiving a JSON object via an HTTP POST request, which includes the content, model, and method to be used. The algorithm first extracts these inputs from the request data. It then configures the environment by setting up the required parameters such as the file containing the bot prompt template, the URL to the API, and the authentication key based on the chosen model and method. The bot prompt file is loaded, and placeholders are replaced with the actual content provided. Next, it prepares the API request by creating authentication headers using the key and formatting the POST data with the model and the customized bot prompt. The algorithm then makes the API call, sending the POST request to the specified URL with the prepared headers and POST data. Once the response is received, it checks for success. If successful, it extracts and formats code blocks from the payload and returns them. If not successful, it handles the error and creates an error response. This process enables dynamic generation of code suited for both backend and frontend development based on the specified method, streamlining the development and testing phases.

we present state-of-the-art work that enables code generation for specific parts of development, including backend, frontend, unit testing, and end-to-end testing, using models such as GPT-3.5, GPT-4, and Llama3. Despite the results, many challenges and research directions exist when developing a modular approach to generate test cases and code. An example of the backend code for the login API in Python, along with its unit test, can be seen in the listing \ref{lst:loginAPI}. 

Similarly, the frontend code generated in the React framework for the login feature can also be found in the listing \ref{lst:loginFrontend}

\begin{lstlisting}[language=Python, caption=Backend Login API, label={lst:loginAPI}]
from flask import Flask, request, jsonify
import unittest
app = Flask(__name__)
users = {
    "user1": "password1",
    "user2": "password2"
}
@app.route('/login', methods=['POST'])
def login():
    data = request.get_json()
    username = data.get('username')
    password = data.get('password')
    if username in users and users[username] == password:
        return jsonify({'message': 'Login successful'})
    else:
        return jsonify({'message': 'Invalid credentials'})
        
// Unit test 
from login_api import LoginAPI
class TestLoginAPI(unittest.TestCase):
    def setUp(self):
        self.login_api = LoginAPI()
    def test_login_success(self):
        response = self.login_api.login(username='test_user', password='password')
        self.assertEqual(response, 'Login successful')
    def test_login_incorrect_password(self):
        response = self.login_api.login(username='test_user', password='wrong_password')
        self.assertEqual(response, 'Incorrect password')
    def test_login_invalid_username(self):
        response = self.login_api.login(username='invalid_user', password='password')
        self.assertEqual(response, 'Invalid username')
    def tearDown(self):
        del self.login_api
\end{lstlisting}

\begin{lstlisting}[language=JavaScript, caption=Frontend Code, label={lst:loginFrontend}] 
// Login.js 
import React, {useState} from 'react'; 
import axios from 'axios'; 

const Login = () => { 
  const [username, setUsername] = useState(''); 
  const [password, setPassword] = useState(''); 
  const handleLogin = async () => { 
    try { 
      const response = await axios.post('https://api.example.com/login', { 
        username, password 
      }); 
      console.log(response.data); // Add logic to handle successful login 
    } catch (error) { 
      console.error('Error logging in: ', error); // Add logic to handle login error 
    } 
  }; 
  return ( 
    <div> 
      <input type="text" placeholder="Username" value={username} onChange={(e) => setUsername(e.target.value)} /> 
      <input type="password" placeholder="Password" value={password} onChange={(e) => setPassword(e.target.value)} /> 
      <button onClick={handleLogin}>Login</button> 
    </div> 
  ); 
}; 

export default Login; 
\end{lstlisting}

Using large language models (LLMs) code generation from requirements include difficulties in understanding context, the need for diverse parsers to differentiate between code and natural language, and a tendency to produce overly generic code. To address these issues, we need to include fine-tuning models for specific development phases, implementing a real-time code review agent, and developing ready-to-deploy software solutions that enhance the effectiveness and efficiency of software development, ensuring robust, cloud-ready deployment solutions.

In examining the challenges and future directions at each stage of the lifecycle, it becomes apparent that certain common challenges need to be addressed across all phases. Addressing these challenges comprehensively can enhance the overall efficiency and success of the lifecycle, paving the way for more robust and sustainable development in future cycles.

\begin{takeawaybox}{Takeaways}
\scriptsize
 \blackcircle{5} \textbf{Automated Code and Test Generation:}: An automated system uses LLMs such as GPT-3.5, GPT-4, and Llama3 to dynamically generate backend and frontend code, along with unit and end-to-end tests, based on JSON input and API interactions, streamlining software development.\\
 \blackcircle{6} \textbf{Associated Challenges and Future Directions}: The system faces challenges in context interpretation and requirement ambiguity; future improvements focus on expanding language support and refining AI models to enhance code accuracy and deployment readiness.
\end{takeawaybox}

\section{Discussion}
\label{Discussion}
The results from our study emphasize the efficacy of LLMs, particularly GPT-3.5, in redefining software engineering processes across various stages—from requirement gathering to deployment. Our findings indicate that GPT-3.5 outperforms other evaluated models, such as GPT-4 and Llama3, in tasks related to software development. This superior performance is particularly notable in the platform's ability to translate user requirements into structured software deliverables, including user stories, UML diagrams, API specifications, and comprehensive test suites. The platform also excels in offering best practice advice, performing security and compliance checks, and suggesting software architect and design optimizations. The results indicate that the incorporation of lifecycle management systems (LLMs) into the software development process can yield notable improvements in the end software products' efficiency and quality.

The findings of this study have significant \textbf{implications} for the field of software engineering, especially about development process automation and optimization. According to the study, software engineers may depend more on AI to handle repetitive and well-defined tasks, freeing up human resources for more complicated and innovative problem-solving activities. This is because GPT-3.5 effectively automates complex jobs. This change may result in decreased development costs, a shorter time to market, and increased productivity. The fact that our platform complies with European security and compliance standards further suggests that LLMs can be successfully customized to satisfy strict legal requirements, making them a practical option for usage in sectors where data security and privacy are critical. The integration of AI in software engineering also opens up new opportunities for developing more robust, secure, and user-centric software solutions.


Although a number of areas call for more research, the current study offers important insights into the potential of LLMs in software engineering. To begin with, broadening the scope of AI models beyond GPT-3.5, GPT-4, and Llama3 may make it easier to find additional models that are more specialized or efficient for particular software engineering challenges. Subsequent studies may examine how to include further AI technologies, such as machine learning models for decision-making and predictive analytics, into software development cycles. Furthermore, to evaluate the long-term effects of AI integration on software quality, maintenance, and scalability, longitudinal studies are required. In conclusion, given the swift advancement of AI technologies, it will be imperative to consistently assess novel models and revise current ones to sustain the efficacy and significance of AI-powered platforms in software engineering. This ongoing research will help ensure that these AI systems remain adaptive to the changing landscapes of software development requirements and technology standards.

\section{Threats to Validity}
\label{ThreatstoValidity}
This section reports the potential threats to the validity of this research and its results, along with mitigation strategies that could help minimize the impacts of the outlined threats based on \cite{easterbrook2008selecting}. The threats are broadly classified across external validity, construct validity, internal validity, and reliability.

\begin{itemize}
    \item \textbf{External Validity}: The generalizability of this study may be limited by the application of GPT-3.5, GPT-4, Llama3, and other APIs to automate software development processes in various real-world domains. Differences in training and model architecture of these models could lead to different outcomes, impacting the practical deployment of AI-driven tools in industries with stringent regulatory requirements such as healthcare and finance. Future research should explore model adaptation across different domains, evaluate integration with diverse technologies, and confirm the practicality of AI in software development through rigorous testing.
    \item \textbf{Construct Validity}: Construct validity concerns how accurately AI models, particularly GPT-3.5 and GPT-4, transform software requirements into functional code and architectural designs. Issues have been observed where AI simplifies or misinterprets complex requirements, potentially compromising the intended functionalities. To mitigate these discrepancies, we have implemented a feedback system at each phase of software development, allowing adjustments to better align theoretical aims with practical outcomes.
    \item \textbf{Internal Validity}: Factors such as project complexity and team expertise with AI can affect the internal validity of our study on multi-agent systems for software development. The stochastic nature of AI models introduces variability, which may lead to misalignments in software development tasks. Addressing these concerns will require rigorous controls, bias mitigation strategies, consistency checks, and comprehensive testing to ensure findings are robust across diverse development scenarios.
    \item \textbf{Reliability}: The reliability of our multi-agent platform is challenged by the inherent variability of LLMs, which can produce different outputs from the same inputs. This variability can impact software functionality and behavior, essential for deployment and maintenance. Strategies to enhance reliability include employing ensemble methods to stabilize model predictions, using strict seed settings to reduce output randomness and further research into model stabilization techniques and consistency checks.
\end{itemize}

\section{Conclusions}
\label{Conclusions}

This study proposed a multiagent unified platform that offers a novel investigation into how LLMs might transform the field of SE. Our work has demonstrated notable progress toward automating the conversion from user requirements to fully realized software deliverables through the development and assessment of a unified AI-driven platform. The platform provides a modular approach to testing and deploying software at different phases of the SDLC. It does this by integrating multiple AI agents, including the most advanced models such as GPT-3.5, GPT-4, and Llama3.

According to the results of our manual investigation, GPT-3.5 performs better than its competitors in the majority of the activities, providing a more dependable and effective method for automating software development chores. This superiority highlights how important it is to select models that are suited for given task needs and development circumstances. In addition to expediting the development process, our uniform platform meets security and compliance requirements, particularly those that correspond with European standards. Furthermore, a major value addition that improves the overall robustness and quality of the software solutions is the platform's capacity to provide design improvements and best practices. 

A move toward more agile, responsive, and user-focused development approaches is promised by the incorporation of LLMs into software engineering techniques. To fully utilize artificial intelligence (AI) in software engineering, this study lays the groundwork for future developments that will see AI not only assist but also take the lead in the development of novel and dependable software solutions.

There are many intriguing prospects for more research because of the way LLM capabilities are constantly evolving and how they are used in software programming. Software development processes will become even more effective and efficient in areas like enhanced model training methods, model modification for certain SE tasks, and the incorporation of more sophisticated AI tools. Future research will continue to improve these tools by pushing the limits of artificial intelligence in this area, making sure they can satisfy the always-changing and expanding needs of the software business.

\section{Acknowledgments}

We express our sincere gratitude to Business Finland for their generous support and funding of our project. Their commitment to fostering innovation and supporting research initiatives has been instrumental in the success of our work.

\bibliographystyle{ACM-Reference-Format}
\bibliography{sample-base}


\begin{thebibliography}{55}


\ifx \showCODEN    \undefined \def \showCODEN     #1{\unskip}     \fi
\ifx \showDOI      \undefined \def \showDOI       #1{#1}\fi
\ifx \showISBNx    \undefined \def \showISBNx     #1{\unskip}     \fi
\ifx \showISBNxiii \undefined \def \showISBNxiii  #1{\unskip}     \fi
\ifx \showISSN     \undefined \def \showISSN      #1{\unskip}     \fi
\ifx \showLCCN     \undefined \def \showLCCN      #1{\unskip}     \fi
\ifx \shownote     \undefined \def \shownote      #1{#1}          \fi
\ifx \showarticletitle \undefined \def \showarticletitle #1{#1}   \fi
\ifx \showURL      \undefined \def \showURL       {\relax}        \fi
\providecommand\bibfield[2]{#2}
\providecommand\bibinfo[2]{#2}
\providecommand\natexlab[1]{#1}
\providecommand\showeprint[2][]{arXiv:#2}

\bibitem[SAF({[n.\,d.]})]%
        {SAFe_WSJF}
 \bibinfo{year}{[n.\,d.]}\natexlab{}.
\newblock \bibinfo{title}{{Weighted Shortest Job First (WSJF) - Scaled Agile Framework}}.
\newblock
\newblock
\urldef\tempurl%
\url{https://scaledagileframework.com/wsjf/}
\showURL{%
\tempurl}
\newblock
\shownote{Accessed: yyyy-mm-dd}.


\bibitem[Abrahamsson et~al\mbox{.}(2017)]%
        {abrahamsson2017agile}
\bibfield{author}{\bibinfo{person}{Pekka Abrahamsson}, \bibinfo{person}{Outi Salo}, \bibinfo{person}{Jussi Ronkainen}, {and} \bibinfo{person}{Juhani Warsta}.} \bibinfo{year}{2017}\natexlab{}.
\newblock \showarticletitle{Agile software development methods: Review and analysis}.
\newblock \bibinfo{journal}{\emph{arXiv preprint arXiv:1709.08439}} (\bibinfo{year}{2017}).
\newblock


\bibitem[Ahmad et~al\mbox{.}(2023)]%
        {ahmad2023towards}
\bibfield{author}{\bibinfo{person}{Aakash Ahmad}, \bibinfo{person}{Muhammad Waseem}, \bibinfo{person}{Peng Liang}, \bibinfo{person}{Mahdi Fahmideh}, \bibinfo{person}{Mst~Shamima Aktar}, {and} \bibinfo{person}{Tommi Mikkonen}.} \bibinfo{year}{2023}\natexlab{}.
\newblock \showarticletitle{Towards human-bot collaborative software architecting with chatgpt}. In \bibinfo{booktitle}{\emph{Proceedings of the 27th International Conference on Evaluation and Assessment in Software Engineering}}. \bibinfo{pages}{279--285}.
\newblock


\bibitem[Ayd{\i}n and Karaarslan(2023)]%
        {aydin2023chatgpt}
\bibfield{author}{\bibinfo{person}{{\"O}mer Ayd{\i}n} {and} \bibinfo{person}{Enis Karaarslan}.} \bibinfo{year}{2023}\natexlab{}.
\newblock \showarticletitle{Is ChatGPT leading generative AI? What is beyond expectations?}
\newblock \bibinfo{journal}{\emph{What is beyond expectations}} (\bibinfo{year}{2023}).
\newblock


\bibitem[Belzner et~al\mbox{.}(2023)]%
        {belzner2023large}
\bibfield{author}{\bibinfo{person}{Lenz Belzner}, \bibinfo{person}{Thomas Gabor}, {and} \bibinfo{person}{Martin Wirsing}.} \bibinfo{year}{2023}\natexlab{}.
\newblock \showarticletitle{Large language model assisted software engineering: prospects, challenges, and a case study}. In \bibinfo{booktitle}{\emph{International Conference on Bridging the Gap between AI and Reality}}. Springer, \bibinfo{pages}{355--374}.
\newblock


\bibitem[Bjarnason et~al\mbox{.}(2016)]%
        {bjarnason2016multi}
\bibfield{author}{\bibinfo{person}{Elizabeth Bjarnason}, \bibinfo{person}{Michael Unterkalmsteiner}, \bibinfo{person}{Markus Borg}, {and} \bibinfo{person}{Emelie Engstr{\"o}m}.} \bibinfo{year}{2016}\natexlab{}.
\newblock \showarticletitle{A multi-case study of agile requirements engineering and the use of test cases as requirements}.
\newblock \bibinfo{journal}{\emph{Information and Software Technology}}  \bibinfo{volume}{77} (\bibinfo{year}{2016}), \bibinfo{pages}{61--79}.
\newblock


\bibitem[Chang et~al\mbox{.}(2024)]%
        {chang2024survey}
\bibfield{author}{\bibinfo{person}{Yupeng Chang}, \bibinfo{person}{Xu Wang}, \bibinfo{person}{Jindong Wang}, \bibinfo{person}{Yuan Wu}, \bibinfo{person}{Linyi Yang}, \bibinfo{person}{Kaijie Zhu}, \bibinfo{person}{Hao Chen}, \bibinfo{person}{Xiaoyuan Yi}, \bibinfo{person}{Cunxiang Wang}, \bibinfo{person}{Yidong Wang}, {et~al\mbox{.}}} \bibinfo{year}{2024}\natexlab{}.
\newblock \showarticletitle{A survey on evaluation of large language models}.
\newblock \bibinfo{journal}{\emph{ACM Transactions on Intelligent Systems and Technology}} \bibinfo{volume}{15}, \bibinfo{number}{3} (\bibinfo{year}{2024}), \bibinfo{pages}{1--45}.
\newblock


\bibitem[Dong et~al\mbox{.}(2023)]%
        {dong2023self}
\bibfield{author}{\bibinfo{person}{Yihong Dong}, \bibinfo{person}{Xue Jiang}, \bibinfo{person}{Zhi Jin}, {and} \bibinfo{person}{Ge Li}.} \bibinfo{year}{2023}\natexlab{}.
\newblock \showarticletitle{Self-collaboration Code Generation via ChatGPT}.
\newblock \bibinfo{journal}{\emph{arXiv preprint arXiv:2304.07590}} (\bibinfo{year}{2023}).
\newblock


\bibitem[Easterbrook et~al\mbox{.}(2008)]%
        {easterbrook2008selecting}
\bibfield{author}{\bibinfo{person}{Steve Easterbrook}, \bibinfo{person}{Janice Singer}, \bibinfo{person}{Margaret-Anne Storey}, {and} \bibinfo{person}{Daniela Damian}.} \bibinfo{year}{2008}\natexlab{}.
\newblock \showarticletitle{Selecting Empirical Methods for Software Engineering Research}.
\newblock In \bibinfo{booktitle}{\emph{Guide to Advanced Empirical Software Engineering}}. \bibinfo{publisher}{Springer}, \bibinfo{pages}{285--311}.
\newblock


\bibitem[Ebert and Louridas(2023)]%
        {ebert2023generative}
\bibfield{author}{\bibinfo{person}{Christof Ebert} {and} \bibinfo{person}{Panos Louridas}.} \bibinfo{year}{2023}\natexlab{}.
\newblock \showarticletitle{Generative AI for software practitioners}.
\newblock \bibinfo{journal}{\emph{IEEE Software}} \bibinfo{volume}{40}, \bibinfo{number}{4} (\bibinfo{year}{2023}), \bibinfo{pages}{30--38}.
\newblock


\bibitem[Feng and Chen(2024)]%
        {feng2024prompting}
\bibfield{author}{\bibinfo{person}{Sidong Feng} {and} \bibinfo{person}{Chunyang Chen}.} \bibinfo{year}{2024}\natexlab{}.
\newblock \showarticletitle{Prompting Is All You Need: Automated Android Bug Replay with Large Language Models}. In \bibinfo{booktitle}{\emph{Proceedings of the 46th IEEE/ACM International Conference on Software Engineering}}. \bibinfo{pages}{1--13}.
\newblock


\bibitem[Feng et~al\mbox{.}(2023)]%
        {feng2023investigating}
\bibfield{author}{\bibinfo{person}{Yunhe Feng}, \bibinfo{person}{Sreecharan Vanam}, \bibinfo{person}{Manasa Cherukupally}, \bibinfo{person}{Weijian Zheng}, \bibinfo{person}{Meikang Qiu}, {and} \bibinfo{person}{Haihua Chen}.} \bibinfo{year}{2023}\natexlab{}.
\newblock \showarticletitle{Investigating Code Generation Performance of Chat-GPT with Crowdsourcing Social Data}. In \bibinfo{booktitle}{\emph{Proceedings of the 47th IEEE Computer Software and Applications Conference}}. \bibinfo{pages}{1--10}.
\newblock


\bibitem[{GPT Laboratory}(2024)]%
        {GPTLaboratory2024}
\bibfield{author}{\bibinfo{person}{{GPT Laboratory}}.} \bibinfo{year}{2024}\natexlab{}.
\newblock \bibinfo{title}{{GPT Development Tool}}.
\newblock \bibinfo{howpublished}{\url{https://github.com/GPT-Laboratory/gpt-developlement-tool}}.
\newblock


\bibitem[Hacker et~al\mbox{.}(2023)]%
        {hacker2023regulating}
\bibfield{author}{\bibinfo{person}{Philipp Hacker}, \bibinfo{person}{Andreas Engel}, {and} \bibinfo{person}{Marco Mauer}.} \bibinfo{year}{2023}\natexlab{}.
\newblock \showarticletitle{Regulating ChatGPT and other large generative AI models}. In \bibinfo{booktitle}{\emph{Proceedings of the 2023 ACM Conference on Fairness, Accountability, and Transparency}}. \bibinfo{pages}{1112--1123}.
\newblock


\bibitem[Huang et~al\mbox{.}(2023)]%
        {huang2023agentcoder}
\bibfield{author}{\bibinfo{person}{Dong Huang}, \bibinfo{person}{Qingwen Bu}, \bibinfo{person}{Jie~M Zhang}, \bibinfo{person}{Michael Luck}, {and} \bibinfo{person}{Heming Cui}.} \bibinfo{year}{2023}\natexlab{}.
\newblock \showarticletitle{AgentCoder: Multi-Agent-based Code Generation with Iterative Testing and Optimisation}.
\newblock \bibinfo{journal}{\emph{arXiv preprint arXiv:2312.13010}} (\bibinfo{year}{2023}).
\newblock


\bibitem[Kirova et~al\mbox{.}(2024)]%
        {kirova2024software}
\bibfield{author}{\bibinfo{person}{Vassilka~D Kirova}, \bibinfo{person}{Cyril~S Ku}, \bibinfo{person}{Joseph~R Laracy}, {and} \bibinfo{person}{Thomas~J Marlowe}.} \bibinfo{year}{2024}\natexlab{}.
\newblock \showarticletitle{Software Engineering Education Must Adapt and Evolve for an LLM Environment}. In \bibinfo{booktitle}{\emph{Proceedings of the 55th ACM Technical Symposium on Computer Science Education V. 1}}. \bibinfo{pages}{666--672}.
\newblock


\bibitem[Li et~al\mbox{.}(2024)]%
        {li2024sheetcopilot}
\bibfield{author}{\bibinfo{person}{Hongxin Li}, \bibinfo{person}{Jingran Su}, \bibinfo{person}{Yuntao Chen}, \bibinfo{person}{Qing Li}, {and} \bibinfo{person}{ZHAO-XIANG ZHANG}.} \bibinfo{year}{2024}\natexlab{}.
\newblock \showarticletitle{SheetCopilot: Bringing Software Productivity to the Next Level through Large Language Models}.
\newblock \bibinfo{journal}{\emph{Advances in Neural Information Processing Systems}}  \bibinfo{volume}{36} (\bibinfo{year}{2024}).
\newblock


\bibitem[Lin et~al\mbox{.}(2024)]%
        {lin2024llm}
\bibfield{author}{\bibinfo{person}{Feng Lin}, \bibinfo{person}{Dong~Jae Kim}, {et~al\mbox{.}}} \bibinfo{year}{2024}\natexlab{}.
\newblock \showarticletitle{When LLM-based Code Generation Meets the Software Development Process}.
\newblock \bibinfo{journal}{\emph{arXiv preprint arXiv:2403.15852}} (\bibinfo{year}{2024}).
\newblock


\bibitem[Liu et~al\mbox{.}(2023)]%
        {liu2023gpt}
\bibfield{author}{\bibinfo{person}{Xiao Liu}, \bibinfo{person}{Yanan Zheng}, \bibinfo{person}{Zhengxiao Du}, \bibinfo{person}{Ming Ding}, \bibinfo{person}{Yujie Qian}, \bibinfo{person}{Zhilin Yang}, {and} \bibinfo{person}{Jie Tang}.} \bibinfo{year}{2023}\natexlab{}.
\newblock \showarticletitle{GPT understands, too}.
\newblock \bibinfo{journal}{\emph{AI Open}} (\bibinfo{year}{2023}).
\newblock


\bibitem[Ma et~al\mbox{.}(2023)]%
        {ma2023scope}
\bibfield{author}{\bibinfo{person}{Wei Ma}, \bibinfo{person}{Shangqing Liu}, \bibinfo{person}{Wenhan Wang}, \bibinfo{person}{Qiang Hu}, \bibinfo{person}{Ye Liu}, \bibinfo{person}{Cen Zhang}, \bibinfo{person}{Liming Nie}, {and} \bibinfo{person}{Yang Liu}.} \bibinfo{year}{2023}\natexlab{}.
\newblock \showarticletitle{The Scope of ChatGPT in Software Engineering: A Thorough Investigation}.
\newblock \bibinfo{journal}{\emph{arXiv preprint arXiv:2305.12138}} (\bibinfo{year}{2023}).
\newblock


\bibitem[MacNeil et~al\mbox{.}(2023)]%
        {macneil2023experiences}
\bibfield{author}{\bibinfo{person}{Stephen MacNeil}, \bibinfo{person}{Andrew Tran}, \bibinfo{person}{Arto Hellas}, \bibinfo{person}{Joanne Kim}, \bibinfo{person}{Sami Sarsa}, \bibinfo{person}{Paul Denny}, \bibinfo{person}{Seth Bernstein}, {and} \bibinfo{person}{Juho Leinonen}.} \bibinfo{year}{2023}\natexlab{}.
\newblock \showarticletitle{Experiences from using code explanations generated by large language models in a web software development e-book}. In \bibinfo{booktitle}{\emph{Proceedings of the 54th ACM Technical Symposium on Computer Science Education V. 1}}. \bibinfo{pages}{931--937}.
\newblock


\bibitem[{Meta Platforms, Inc.}(2024)]%
        {metallama}
\bibfield{author}{\bibinfo{person}{{Meta Platforms, Inc.}}} \bibinfo{year}{2024}\natexlab{}.
\newblock \bibinfo{title}{Meta Llama}.
\newblock \bibinfo{howpublished}{Available at \url{https://llama.meta.com/}}.
\newblock
\newblock
\shownote{Accessed: yyyy-mm-dd}.


\bibitem[Mnkandla and Dwolatzky(2004)]%
        {9487986}
\bibfield{author}{\bibinfo{person}{E. Mnkandla} {and} \bibinfo{person}{B. Dwolatzky}.} \bibinfo{year}{2004}\natexlab{}.
\newblock \showarticletitle{A survey of agile methodologies}.
\newblock \bibinfo{journal}{\emph{Transactions of the South African Institute of Electrical Engineers}} \bibinfo{volume}{95}, \bibinfo{number}{4} (\bibinfo{year}{2004}), \bibinfo{pages}{236--247}.
\newblock


\bibitem[Mucha et~al\mbox{.}(2024)]%
        {mucha2024systematic}
\bibfield{author}{\bibinfo{person}{Julia Mucha}, \bibinfo{person}{Andreas Kaufmann}, {and} \bibinfo{person}{Dirk Riehle}.} \bibinfo{year}{2024}\natexlab{}.
\newblock \showarticletitle{A systematic literature review of pre-requirements specification traceability}.
\newblock \bibinfo{journal}{\emph{Requirements Engineering}} (\bibinfo{year}{2024}), \bibinfo{pages}{1--23}.
\newblock


\bibitem[Nguyen-Duc et~al\mbox{.}(2023)]%
        {nguyen2023generative}
\bibfield{author}{\bibinfo{person}{Anh Nguyen-Duc}, \bibinfo{person}{Beatriz Cabrero-Daniel}, \bibinfo{person}{Adam Przybylek}, \bibinfo{person}{Chetan Arora}, \bibinfo{person}{Dron Khanna}, \bibinfo{person}{Tomas Herda}, \bibinfo{person}{Usman Rafiq}, \bibinfo{person}{Jorge Melegati}, \bibinfo{person}{Eduardo Guerra}, \bibinfo{person}{Kai-Kristian Kemell}, {et~al\mbox{.}}} \bibinfo{year}{2023}\natexlab{}.
\newblock \showarticletitle{Generative Artificial Intelligence for Software Engineering--A Research Agenda}.
\newblock \bibinfo{journal}{\emph{arXiv preprint arXiv:2310.18648}} (\bibinfo{year}{2023}).
\newblock


\bibitem[Ouyang et~al\mbox{.}(2022)]%
        {ouyang2022training}
\bibfield{author}{\bibinfo{person}{Long Ouyang}, \bibinfo{person}{Jeffrey Wu}, \bibinfo{person}{Xu Jiang}, \bibinfo{person}{Diogo Almeida}, \bibinfo{person}{Carroll Wainwright}, \bibinfo{person}{Pamela Mishkin}, \bibinfo{person}{Chong Zhang}, \bibinfo{person}{Sandhini Agarwal}, \bibinfo{person}{Katarina Slama}, \bibinfo{person}{Alex Ray}, {et~al\mbox{.}}} \bibinfo{year}{2022}\natexlab{}.
\newblock \showarticletitle{Training language models to follow instructions with human feedback}.
\newblock \bibinfo{journal}{\emph{Advances in Neural Information Processing Systems}}  \bibinfo{volume}{35} (\bibinfo{year}{2022}), \bibinfo{pages}{27730--27744}.
\newblock


\bibitem[{PlantUML Team}(2024)]%
        {PlantUML}
\bibfield{author}{\bibinfo{person}{{PlantUML Team}}.} \bibinfo{year}{2024}\natexlab{}.
\newblock \bibinfo{title}{PlantUML}.
\newblock
\newblock
\urldef\tempurl%
\url{http://plantuml.com}
\showURL{%
\tempurl}
\newblock
\shownote{Accessed: 2024-05-01}.


\bibitem[Qian et~al\mbox{.}(2023)]%
        {qian2023communicative}
\bibfield{author}{\bibinfo{person}{Chen Qian}, \bibinfo{person}{Xin Cong}, \bibinfo{person}{Cheng Yang}, \bibinfo{person}{Weize Chen}, \bibinfo{person}{Yusheng Su}, \bibinfo{person}{Juyuan Xu}, \bibinfo{person}{Zhiyuan Liu}, {and} \bibinfo{person}{Maosong Sun}.} \bibinfo{year}{2023}\natexlab{}.
\newblock \showarticletitle{Communicative agents for software development}.
\newblock \bibinfo{journal}{\emph{arXiv preprint arXiv:2307.07924}} (\bibinfo{year}{2023}).
\newblock


\bibitem[Quevedo et~al\mbox{.}(2024)]%
        {quevedo2024evaluating}
\bibfield{author}{\bibinfo{person}{Ernesto Quevedo}, \bibinfo{person}{Amr~S Abdelfattah}, \bibinfo{person}{Alejandro Rodriguez}, \bibinfo{person}{Jorge Yero}, {and} \bibinfo{person}{Tomas Cerny}.} \bibinfo{year}{2024}\natexlab{}.
\newblock \showarticletitle{Evaluating ChatGPT’s Proficiency in Understanding and Answering Microservice Architecture Queries Using Source Code Insights}.
\newblock \bibinfo{journal}{\emph{SN Computer Science}} \bibinfo{volume}{5}, \bibinfo{number}{4} (\bibinfo{year}{2024}), \bibinfo{pages}{422}.
\newblock


\bibitem[Radford et~al\mbox{.}(2018)]%
        {radford2018improving}
\bibfield{author}{\bibinfo{person}{Alec Radford}, \bibinfo{person}{Karthik Narasimhan}, \bibinfo{person}{Tim Salimans}, \bibinfo{person}{Ilya Sutskever}, {et~al\mbox{.}}} \bibinfo{year}{2018}\natexlab{}.
\newblock \showarticletitle{Improving language understanding by generative pre-training}.
\newblock  (\bibinfo{year}{2018}).
\newblock


\bibitem[Rajbhoj et~al\mbox{.}(2024)]%
        {rajbhoj2024accelerating}
\bibfield{author}{\bibinfo{person}{Asha Rajbhoj}, \bibinfo{person}{Akanksha Somase}, \bibinfo{person}{Piyush Kulkarni}, {and} \bibinfo{person}{Vinay Kulkarni}.} \bibinfo{year}{2024}\natexlab{}.
\newblock \showarticletitle{Accelerating Software Development Using Generative AI: ChatGPT Case Study}. In \bibinfo{booktitle}{\emph{Proceedings of the 17th Innovations in Software Engineering Conference}}. \bibinfo{pages}{1--11}.
\newblock


\bibitem[Rasheed et~al\mbox{.}(2024a)]%
        {rasheed2024can}
\bibfield{author}{\bibinfo{person}{Zeeshan Rasheed}, \bibinfo{person}{Muhammad Waseem}, \bibinfo{person}{Aakash Ahmad}, \bibinfo{person}{Kai-Kristian Kemell}, \bibinfo{person}{Wang Xiaofeng}, \bibinfo{person}{Anh~Nguyen Duc}, {and} \bibinfo{person}{Pekka Abrahamsson}.} \bibinfo{year}{2024}\natexlab{a}.
\newblock \showarticletitle{Can Large Language Models Serve as Data Analysts? A Multi-Agent Assisted Approach for Qualitative Data Analysis}.
\newblock \bibinfo{journal}{\emph{arXiv preprint arXiv:2402.01386}} (\bibinfo{year}{2024}).
\newblock


\bibitem[Rasheed et~al\mbox{.}(2023)]%
        {rasheed2023autonomous}
\bibfield{author}{\bibinfo{person}{Zeeshan Rasheed}, \bibinfo{person}{Muhammad Waseem}, \bibinfo{person}{Kai-Kristian Kemell}, \bibinfo{person}{Wang Xiaofeng}, \bibinfo{person}{Anh~Nguyen Duc}, \bibinfo{person}{Kari Syst{\"a}}, {and} \bibinfo{person}{Pekka Abrahamsson}.} \bibinfo{year}{2023}\natexlab{}.
\newblock \showarticletitle{Autonomous Agents in Software Development: A Vision Paper}.
\newblock \bibinfo{journal}{\emph{arXiv preprint arXiv:2311.18440}} (\bibinfo{year}{2023}).
\newblock


\bibitem[Rasheed et~al\mbox{.}(2024b)]%
        {rasheed2024codepori}
\bibfield{author}{\bibinfo{person}{Zeeshan Rasheed}, \bibinfo{person}{Muhammad Waseem}, \bibinfo{person}{Mika Saari}, \bibinfo{person}{Kari Syst{\"a}}, {and} \bibinfo{person}{Pekka Abrahamsson}.} \bibinfo{year}{2024}\natexlab{b}.
\newblock \showarticletitle{Codepori: Large scale model for autonomous software development by using multi-agents}.
\newblock \bibinfo{journal}{\emph{arXiv preprint arXiv:2402.01411}} (\bibinfo{year}{2024}).
\newblock


\bibitem[Rasheed et~al\mbox{.}({[n.\,d.]})]%
        {rasheed2024large}
\bibfield{author}{\bibinfo{person}{Zeeshan Rasheed}, \bibinfo{person}{Muhammad Waseem}, \bibinfo{person}{Kari Syst{\"a}}, {and} \bibinfo{person}{Pekka Abrahamsson}.} \bibinfo{year}{[n.\,d.]}\natexlab{}.
\newblock \showarticletitle{Large Language Model Evaluation Via Multi AI Agents: Preliminary results}. In \bibinfo{booktitle}{\emph{ICLR 2024 Workshop on Large Language Model (LLM) Agents}}.
\newblock


\bibitem[Rinard(2024)]%
        {rinard2024software}
\bibfield{author}{\bibinfo{person}{Martin Rinard}.} \bibinfo{year}{2024}\natexlab{}.
\newblock \showarticletitle{Software Engineering Research in a World with Generative Artificial Intelligence}. In \bibinfo{booktitle}{\emph{2024 IEEE/ACM 46th International Conference on Software Engineering (ICSE)}}. IEEE Computer Society, \bibinfo{pages}{3--7}.
\newblock


\bibitem[Roumeliotis and Tselikas(2023)]%
        {roumeliotis2023chatgpt}
\bibfield{author}{\bibinfo{person}{Konstantinos~I Roumeliotis} {and} \bibinfo{person}{Nikolaos~D Tselikas}.} \bibinfo{year}{2023}\natexlab{}.
\newblock \showarticletitle{Chatgpt and open-ai models: A preliminary review}.
\newblock \bibinfo{journal}{\emph{Future Internet}} \bibinfo{volume}{15}, \bibinfo{number}{6} (\bibinfo{year}{2023}), \bibinfo{pages}{192}.
\newblock


\bibitem[Sami et~al\mbox{.}(2024a)]%
        {sami2024system}
\bibfield{author}{\bibinfo{person}{Abdul~Malik Sami}, \bibinfo{person}{Zeeshan Rasheed}, \bibinfo{person}{Kai-Kristian Kemell}, \bibinfo{person}{Muhammad Waseem}, \bibinfo{person}{Terhi Kilamo}, \bibinfo{person}{Mika Saari}, \bibinfo{person}{Anh~Nguyen Duc}, \bibinfo{person}{Kari Syst{\"a}}, {and} \bibinfo{person}{Pekka Abrahamsson}.} \bibinfo{year}{2024}\natexlab{a}.
\newblock \showarticletitle{System for systematic literature review using multiple AI agents: Concept and an empirical evaluation}.
\newblock \bibinfo{journal}{\emph{arXiv preprint arXiv:2403.08399}} (\bibinfo{year}{2024}).
\newblock


\bibitem[Sami et~al\mbox{.}({[n.\,d.]})]%
        {samiprioritizing}
\bibfield{author}{\bibinfo{person}{Malik~Abdul Sami}, \bibinfo{person}{Zeeshan Rasheed}, \bibinfo{person}{Muhammad Waseem}, \bibinfo{person}{Zheying Zhang}, \bibinfo{person}{Tomas Herda}, {and} \bibinfo{person}{Pekka Abrahamsson}.} \bibinfo{year}{[n.\,d.]}\natexlab{}.
\newblock \showarticletitle{Prioritizing Software Requirements Using Large Language Models}.
\newblock  (\bibinfo{year}{[n.\,d.]}).
\newblock


\bibitem[Sami et~al\mbox{.}(2024b)]%
        {sami2024prioritizing}
\bibfield{author}{\bibinfo{person}{Malik~Abdul Sami}, \bibinfo{person}{Zeeshan Rasheed}, \bibinfo{person}{Muhammad Waseem}, \bibinfo{person}{Zheying Zhang}, \bibinfo{person}{Tomas Herda}, {and} \bibinfo{person}{Pekka Abrahamsson}.} \bibinfo{year}{2024}\natexlab{b}.
\newblock \bibinfo{title}{Prioritizing Software Requirements Using Large Language Models}.
\newblock
\newblock
\showeprint[arxiv]{2405.01564}~[cs.SE]


\bibitem[Sauvola et~al\mbox{.}(2024)]%
        {sauvola2024future}
\bibfield{author}{\bibinfo{person}{Jaakko Sauvola}, \bibinfo{person}{Sasu Tarkoma}, \bibinfo{person}{Mika Klemettinen}, \bibinfo{person}{Jukka Riekki}, {and} \bibinfo{person}{David Doermann}.} \bibinfo{year}{2024}\natexlab{}.
\newblock \showarticletitle{Future of software development with generative AI}.
\newblock \bibinfo{journal}{\emph{Automated Software Engineering}} \bibinfo{volume}{31}, \bibinfo{number}{1} (\bibinfo{year}{2024}), \bibinfo{pages}{26}.
\newblock


\bibitem[Svensson and Torkar(2024)]%
        {svensson2024not}
\bibfield{author}{\bibinfo{person}{Richard~Berntsson Svensson} {and} \bibinfo{person}{Richard Torkar}.} \bibinfo{year}{2024}\natexlab{}.
\newblock \showarticletitle{Not all requirements prioritization criteria are equal at all times: A quantitative analysis}.
\newblock \bibinfo{journal}{\emph{Journal of Systems and Software}}  \bibinfo{volume}{209} (\bibinfo{year}{2024}), \bibinfo{pages}{111909}.
\newblock


\bibitem[Tang et~al\mbox{.}(2024a)]%
        {tang2024collaborative}
\bibfield{author}{\bibinfo{person}{Daniel Tang}, \bibinfo{person}{Zhenghan Chen}, \bibinfo{person}{Kisub Kim}, \bibinfo{person}{Yewei Song}, \bibinfo{person}{Haoye Tian}, \bibinfo{person}{Saad Ezzini}, \bibinfo{person}{Yongfeng Huang}, {and} \bibinfo{person}{Jacques Klein Tegawende~F Bissyande}.} \bibinfo{year}{2024}\natexlab{a}.
\newblock \showarticletitle{Collaborative agents for software engineering}.
\newblock \bibinfo{journal}{\emph{arXiv preprint arXiv:2402.02172}} (\bibinfo{year}{2024}).
\newblock


\bibitem[Tang et~al\mbox{.}(2024b)]%
        {tang2024chatgpt}
\bibfield{author}{\bibinfo{person}{Yutian Tang}, \bibinfo{person}{Zhijie Liu}, \bibinfo{person}{Zhichao Zhou}, {and} \bibinfo{person}{Xiapu Luo}.} \bibinfo{year}{2024}\natexlab{b}.
\newblock \showarticletitle{Chatgpt vs sbst: A comparative assessment of unit test suite generation}.
\newblock \bibinfo{journal}{\emph{IEEE Transactions on Software Engineering}} (\bibinfo{year}{2024}).
\newblock


\bibitem[Tikayat~Ray et~al\mbox{.}(2023)]%
        {tikayat2023agile}
\bibfield{author}{\bibinfo{person}{Archana Tikayat~Ray}, \bibinfo{person}{Bjorn~F Cole}, \bibinfo{person}{Olivia~J Pinon~Fischer}, \bibinfo{person}{Anirudh~Prabhakara Bhat}, \bibinfo{person}{Ryan~T White}, {and} \bibinfo{person}{Dimitri~N Mavris}.} \bibinfo{year}{2023}\natexlab{}.
\newblock \showarticletitle{Agile Methodology for the Standardization of Engineering Requirements Using Large Language Models}.
\newblock \bibinfo{journal}{\emph{Systems}} \bibinfo{volume}{11}, \bibinfo{number}{7} (\bibinfo{year}{2023}), \bibinfo{pages}{352}.
\newblock


\bibitem[Touvron et~al\mbox{.}(2023)]%
        {touvron2023llama}
\bibfield{author}{\bibinfo{person}{Hugo Touvron}, \bibinfo{person}{Thibaut Lavril}, \bibinfo{person}{Gautier Izacard}, \bibinfo{person}{Xavier Martinet}, \bibinfo{person}{Marie-Anne Lachaux}, \bibinfo{person}{Timoth{\'e}e Lacroix}, \bibinfo{person}{Baptiste Rozi{\`e}re}, \bibinfo{person}{Naman Goyal}, \bibinfo{person}{Eric Hambro}, \bibinfo{person}{Faisal Azhar}, {et~al\mbox{.}}} \bibinfo{year}{2023}\natexlab{}.
\newblock \showarticletitle{Llama: Open and efficient foundation language models}.
\newblock \bibinfo{journal}{\emph{arXiv preprint arXiv:2302.13971}} (\bibinfo{year}{2023}).
\newblock


\bibitem[Treude(2023)]%
        {treude2023navigating}
\bibfield{author}{\bibinfo{person}{Christoph Treude}.} \bibinfo{year}{2023}\natexlab{}.
\newblock \showarticletitle{Navigating Complexity in Software Engineering: A Prototype for Comparing GPT-n Solutions}.
\newblock \bibinfo{journal}{\emph{arXiv preprint arXiv:2301.12169}} (\bibinfo{year}{2023}).
\newblock


\bibitem[Wang et~al\mbox{.}(2024a)]%
        {wang2024software}
\bibfield{author}{\bibinfo{person}{Junjie Wang}, \bibinfo{person}{Yuchao Huang}, \bibinfo{person}{Chunyang Chen}, \bibinfo{person}{Zhe Liu}, \bibinfo{person}{Song Wang}, {and} \bibinfo{person}{Qing Wang}.} \bibinfo{year}{2024}\natexlab{a}.
\newblock \showarticletitle{Software testing with large language models: Survey, landscape, and vision}.
\newblock \bibinfo{journal}{\emph{IEEE Transactions on Software Engineering}} (\bibinfo{year}{2024}).
\newblock


\bibitem[Wang et~al\mbox{.}(2024b)]%
        {wang2024xuat}
\bibfield{author}{\bibinfo{person}{Zhitao Wang}, \bibinfo{person}{Wei Wang}, \bibinfo{person}{Zirao Li}, \bibinfo{person}{Long Wang}, \bibinfo{person}{Can Yi}, \bibinfo{person}{Xinjie Xu}, \bibinfo{person}{Luyang Cao}, \bibinfo{person}{Hanjing Su}, \bibinfo{person}{Shouzhi Chen}, {and} \bibinfo{person}{Jun Zhou}.} \bibinfo{year}{2024}\natexlab{b}.
\newblock \showarticletitle{XUAT-Copilot: Multi-Agent Collaborative System for Automated User Acceptance Testing with Large Language Model}.
\newblock \bibinfo{journal}{\emph{arXiv preprint arXiv:2401.02705}} (\bibinfo{year}{2024}).
\newblock


\bibitem[Waseem et~al\mbox{.}(2023a)]%
        {waseem2023using}
\bibfield{author}{\bibinfo{person}{Muhammad Waseem}, \bibinfo{person}{Teerath Das}, \bibinfo{person}{Aakash Ahmad}, \bibinfo{person}{Mahdi Fehmideh}, \bibinfo{person}{Peng Liang}, {and} \bibinfo{person}{Tommi Mikkonen}.} \bibinfo{year}{2023}\natexlab{a}.
\newblock \showarticletitle{Using ChatGPT throughout the Software Development Life Cycle by Novice Developers}.
\newblock \bibinfo{journal}{\emph{arXiv preprint arXiv:2310.13648}} (\bibinfo{year}{2023}).
\newblock


\bibitem[Waseem et~al\mbox{.}(2023b)]%
        {waseem2023artificial}
\bibfield{author}{\bibinfo{person}{Muhammad Waseem}, \bibinfo{person}{Teerath Das}, \bibinfo{person}{Teemu Paloniemi}, \bibinfo{person}{Miika Koivisto}, \bibinfo{person}{Eeli R{\"a}s{\"a}nen}, \bibinfo{person}{Manu Set{\"a}l{\"a}}, {and} \bibinfo{person}{Tommi Mikkonen}.} \bibinfo{year}{2023}\natexlab{b}.
\newblock \showarticletitle{Artificial Intelligence Procurement Assistant: Enhancing Bid Evaluation}. In \bibinfo{booktitle}{\emph{International Conference on Software Business}}. Springer Nature Switzerland Cham, \bibinfo{pages}{108--114}.
\newblock


\bibitem[Xie et~al\mbox{.}(2023)]%
        {xie2023chatunitest}
\bibfield{author}{\bibinfo{person}{Zhuokui Xie}, \bibinfo{person}{Yinghao Chen}, \bibinfo{person}{Chen Zhi}, \bibinfo{person}{Shuiguang Deng}, {and} \bibinfo{person}{Jianwei Yin}.} \bibinfo{year}{2023}\natexlab{}.
\newblock \showarticletitle{ChatUniTest: a ChatGPT-based automated unit test generation tool}.
\newblock \bibinfo{journal}{\emph{arXiv preprint arXiv:2305.04764}} (\bibinfo{year}{2023}).
\newblock


\bibitem[Zhang et~al\mbox{.}(2024)]%
        {zhang2024llm}
\bibfield{author}{\bibinfo{person}{Zheying Zhang}, \bibinfo{person}{Maruf Rayhan}, \bibinfo{person}{Tomas Herda}, \bibinfo{person}{Manuel Goisauf}, {and} \bibinfo{person}{Pekka Abrahamsson}.} \bibinfo{year}{2024}\natexlab{}.
\newblock \showarticletitle{LLM-based agents for automating the enhancement of user story quality: An early report}.
\newblock \bibinfo{journal}{\emph{arXiv preprint arXiv:2403.09442}} (\bibinfo{year}{2024}).
\newblock


\bibitem[Zhong et~al\mbox{.}(2024)]%
        {zhong2024ldb}
\bibfield{author}{\bibinfo{person}{Li Zhong}, \bibinfo{person}{Zilong Wang}, {and} \bibinfo{person}{Jingbo Shang}.} \bibinfo{year}{2024}\natexlab{}.
\newblock \showarticletitle{LDB: A Large Language Model Debugger via Verifying Runtime Execution Step-by-step}.
\newblock \bibinfo{journal}{\emph{arXiv preprint arXiv:2402.16906}} (\bibinfo{year}{2024}).
\newblock


\bibitem[Zhou et~al\mbox{.}(2023)]%
        {zhou2023language}
\bibfield{author}{\bibinfo{person}{Andy Zhou}, \bibinfo{person}{Kai Yan}, \bibinfo{person}{Michal Shlapentokh-Rothman}, \bibinfo{person}{Haohan Wang}, {and} \bibinfo{person}{Yu-Xiong Wang}.} \bibinfo{year}{2023}\natexlab{}.
\newblock \showarticletitle{Language agent tree search unifies reasoning acting and planning in language models}.
\newblock \bibinfo{journal}{\emph{arXiv preprint arXiv:2310.04406}} (\bibinfo{year}{2023}).
\newblock


\end{thebibliography}


\end{document}